\documentclass[twocolumn,aip,reprint]{revtex4-1}
\usepackage[T1]{fontenc}
\usepackage[latin9]{inputenc}
\usepackage{array}
\usepackage{units}
\usepackage{bm}
\usepackage{multirow}
\usepackage{amsbsy}
\usepackage{graphicx}
\usepackage{subscript}
\makeatletter

\providecommand{\tabularnewline}{\\}
\newcommand{\lyxdot}{.}

\usepackage{times}

\makeatother

\begin{document}

\title{Optimizing the tight-binding parametrization of\\
the quasi-one-dimensional superconductor K\textsubscript{2}Cr\textsubscript{3}As\textsubscript{3}}

\author{Giuseppe Cuono}

\affiliation{Dip. di Fisica ``E.R. Caianiello'', Univ. di Salerno, I-84084 Fisciano
(SA), Italy}

\author{Carmine Autieri}

\affiliation{CNR-SPIN, UOS L'Aquila, Sede Temporanea di Chieti, I-66100 Chieti,
Italy}

\author{Filomena Forte}

\affiliation{Dip. di Fisica ``E.R. Caianiello'', Univ. di Salerno, I-84084 Fisciano
(SA), Italy}

\affiliation{CNR-SPIN, UOS Salerno, I-84084 Fisciano (SA), Italy}

\author{Gaetano Busiello}

\affiliation{Dip. di Fisica ``E.R. Caianiello'', Univ. di Salerno, I-84084 Fisciano
(SA), Italy}

\author{Maria Teresa Mercaldo}

\affiliation{Dip. di Fisica ``E.R. Caianiello'', Univ. di Salerno, I-84084 Fisciano
(SA), Italy}

\author{Alfonso Romano}

\affiliation{Dip. di Fisica ``E.R. Caianiello'', Univ. di Salerno, I-84084 Fisciano
(SA), Italy}

\affiliation{CNR-SPIN, UOS Salerno, I-84084 Fisciano (SA), Italy}

\author{Canio Noce}

\affiliation{Dip. di Fisica ``E.R. Caianiello'', Univ. di Salerno, I-84084 Fisciano
(SA), Italy}

\affiliation{CNR-SPIN, UOS Salerno, I-84084 Fisciano (SA), Italy}

\author{Adolfo Avella}

\thanks{Corresponding author: Adolfo Avella (avella@sa.infn.it)}

\affiliation{Dip. di Fisica ``E.R. Caianiello'', Univ. di Salerno, I-84084 Fisciano
(SA), Italy}

\affiliation{CNR-SPIN, UOS Salerno, I-84084 Fisciano (SA), Italy}

\affiliation{Unit\'{a} CNISM di Salerno, Univ. di Salerno, I-84084 Fisciano (SA),
Italy}

\date{\today}
\begin{abstract}
We study the tight-binding dispersion of the recently discovered superconductor
K\textsubscript{2}Cr\textsubscript{3}As\textsubscript{3}, obtained
from Wannier projection of \textit{\emph{Density Functional Theory
(DFT)}} results. In order to establish quantitatively the actual degree
of quasi-one-dimensionality of this compound, we analyze the electronic
band structure for two reduced sets of hopping parameters: one restricted
to the Cr-As tubes and another one retaining a minimal number of in-plane
hoppings. The corresponding total and local density of states of the
compound are also computed with the aim of assessing the tight-binding
results with respect to the DFT ones. We find a quite good agreement
with the DFT results for the more extended set of hopping parameters,
especially for what concerns the orbitals that dominate at the Fermi
level. Accordingly, we conclude that one cannot avoid taking into
account in-plane hoppings up to the next-nearest-neighbors cells even
only to describe correctly the Fermi surface cuts and the populations
along the $k_{z}$ direction. Such a choice of a minimal number of
hopping parameters directly reflects in the possibility of correctly
describing correlations and magnetic interactions.
\end{abstract}
\maketitle

\section{Introduction}

Systems such as heavy fermion compounds~\citep{Stewart_84,Amato_97,Movshovich01},
high transition-temperature cuprate superconductors~\citep{VanHarlingen95,Timusk_99,Norman_05,Lee_06,Armitage_10,Rice_12,Avella_14a,Keimer_15,Avella_16,Novelli_17,Avella_18},
strontium ruthenate superconductors~\citep{Mackenzie03,Ovchinnikov_03,Cuoco_06,Cuoco_06a,Forte_10,Autieri_12,Malvestuto_13,Autieri_14,Granata_16}
and iron-pnictide superconductors~\citep{Mazin08,Stewart_11} have
been investigated very thoroughly due to their unconventional properties
determined by the existence of a superconductive phase that appears
close to a magnetic one when the latter is suppressed by applying
an \emph{external} agent, such as for instance pressure. In recent
years, this type of investigation has been extended to the search
of superconductivity in Cr pnictides, although chromium is one of
the few metallic elements that doesn't superconduct even under high
pressures~\citep{Bao15}. Accordingly, the discovery of superconductivity
in chromium arsenide (CrAs) was only recent and rather unexpected.
CrAs exhibits a superconducting transition of $T_{c}\approx2$~K
for a critical pressure of $P_{c}\approx8$~Kbar, where its magnetically
ordered phase is completely suppressed~\citep{Wu10,Wu14,Kotegawa14}.
Then, another family of superconducting CrAs-based compounds has been
discovered, that of A\textsubscript{2}Cr\textsubscript{3}As\textsubscript{3},
where A is K~\citep{Bao15}, Rb~\citep{Tang15} or Cs~\citep{Tang15a}.

It is worth noting that while CrAs is a 3D compound and superconducts
only under a sizable pressure, A\textsubscript{2}Cr\textsubscript{3}As\textsubscript{3}
are considered quasi-1D superconductors at ambient pressure. In particular,
K\textsubscript{2}Cr\textsubscript{3}As\textsubscript{3} has a
$T_{C}\approx6.1$~K, an hexagonal crystal structure at room temperature
with $a$=9.9832 {\AA} and $c$=4.2304 {\AA}, and {[}(Cr\textsubscript{3}As\textsubscript{3})\textsuperscript{2-}{]}\textsubscript{$\infty$}
double-walled subnanotubes separated by columns of K\textsuperscript{+}
ions~\citep{Bao15}. This material is very peculiar because it is
considered quasi-one dimensional and, at the same time, it is a superconductor:
superconductivity is not so frequent in quasi-one dimensional compounds
due to the Peierls instability~\citep{Peierls55,Smaalen05}. As regards
K\textsubscript{2}Cr\textsubscript{3}As\textsubscript{3}, a large
specific heat coefficient of 70-75 mJ K\textsuperscript{-2} mol\textsuperscript{-1}
indicates strong electron correlations, while a linear temperature
dependence of the resistivity from 7 to 300 K supports the hypothesis
of a Tomonaga-Luttinger liquid behavior~\citep{Bao15}, as foreseeable
for fermions confined in 1D. On the other hand, Kong \textit{et al.}
report a T\textsuperscript{3} dependence of resistivity from 10 to
40 K~\citep{Kong15} and therefore the degree of unidimensionality
of the system is still under current debate.

All experiments point to an unconventional type of superconductivity:
an upper critical field that is 3-4 times the Pauli limit~\citep{Bao15},
the absence of the Hebel-Schlichter coherence peak in $1/T_{1}$ just
below $T_{C}$~\citep{Zhi15} and a linear temperature dependence
of the penetration depth for $T\ll T_{c}$~\citep{Pang15}. Muon
spin rotation measurements show results consistent with a $d$-wave
model with line nodes. The weak evidence of a spontaneous appearance
of an internal magnetic field below the transition temperature is
probably another proof that the superconducting state is not conventional~\citep{Adroja15}.
The Debye temperature is found to be 220 K, the specific heat jump
at the superconducting transition is 2.2$\gamma T_{c}$ and the upper
critical field is anisotropic with different amplitudes between fields
parallel and perpendicular to the rodlike crystals~\citep{Kong15}.
NMR measurements on the compound show evidence for a strong enhancement
of Cr spin fluctuations above $T_{c}$ in the subnanotubes based on
the nuclear spin-lattice relaxation rate $1/T_{1}$~\citep{Zhi15}.

From a theoretical point of view, first-principles calculations indicate
that the ground state is paramagnetic and that the Cr-$d_{xy},d_{x^{2}-y^{2}},d_{z^{2}}$
orbitals dominate close to the Fermi level, with three bands crossing
$E_{F}$ to form one 3D Fermi surface sheet and two quasi 1D-sheets~\citep{Jiang15}.
Wu \textit{et al} predict that both K\textsubscript{2}Cr\textsubscript{3}As\textsubscript{3}
and Rb\textsubscript{2}Cr\textsubscript{3}As\textsubscript{3} exhibit
strong frustrated magnetic fluctuations and are close to a novel in-out
coplanar magnetic ground state~\citep{Xian-Xin15}. They also report
that frustrated magnetism is very sensitive to the $c$-axis lattice
constant and can thus be suppressed by increasing pressure~\citep{Xian-Xin15}.
Moreover, three- and six- band models have been built and they show
in both weak and strong coupling limits that a triplet $p_{z}$-wave
pairing is the leading pairing symmetry for physically realistic parameters~\citep{Wu15,Zhang16}.
Zhou \textit{et al} have obtained, for A\textsubscript{2}Cr\textsubscript{3}As\textsubscript{3}
(A=K, Rb or Cs), that at small $U$ the pairing comes from 3D $\gamma$
band and has spatial symmetry $f_{y(3x^{2}-y^{2})}$, with line nodes
in the gap function, while at large $U$ a fully gapped $p$-wave
state, $p_{z}\hat{z}$ dominates at the quasi-1D $\alpha$-band~\citep{Yi17}. 

In this paper, we study the electronic properties of K\textsubscript{2}Cr\textsubscript{3}As\textsubscript{3}
by implementing a tight-binding reduction procedure onto state-of-the-art
DFT calculations in order to contribute to clarify the issue of the
quasi-one-dimensionality of the compound. We obtain the band structure
and the total and local density of states (DOS) considering two reduced
sets of hopping parameters and compare them with the DFT results to
establish the minimal number of in-plane hopping amplitudes necessary
to get a reasonable agreement and an adequate description of the properties
of the material. The hopping parameters have been extracted by means
of the Wannier procedure applied to the DFT results. The paper is
organized as follows: in the next section, we describe our tight-binding
procedure and show the band structure obtained considering the two
reduced sets of hopping parameters. In Sec. III, we focus on the total
and the local DOS to check the reliability of the results from the
previous section with respect to the DFT ones. In Sec. IV, we draw
our conclusions and make some final remarks.

\begin{figure}[t]
\noindent \centering{}\includegraphics[width=0.75\columnwidth]{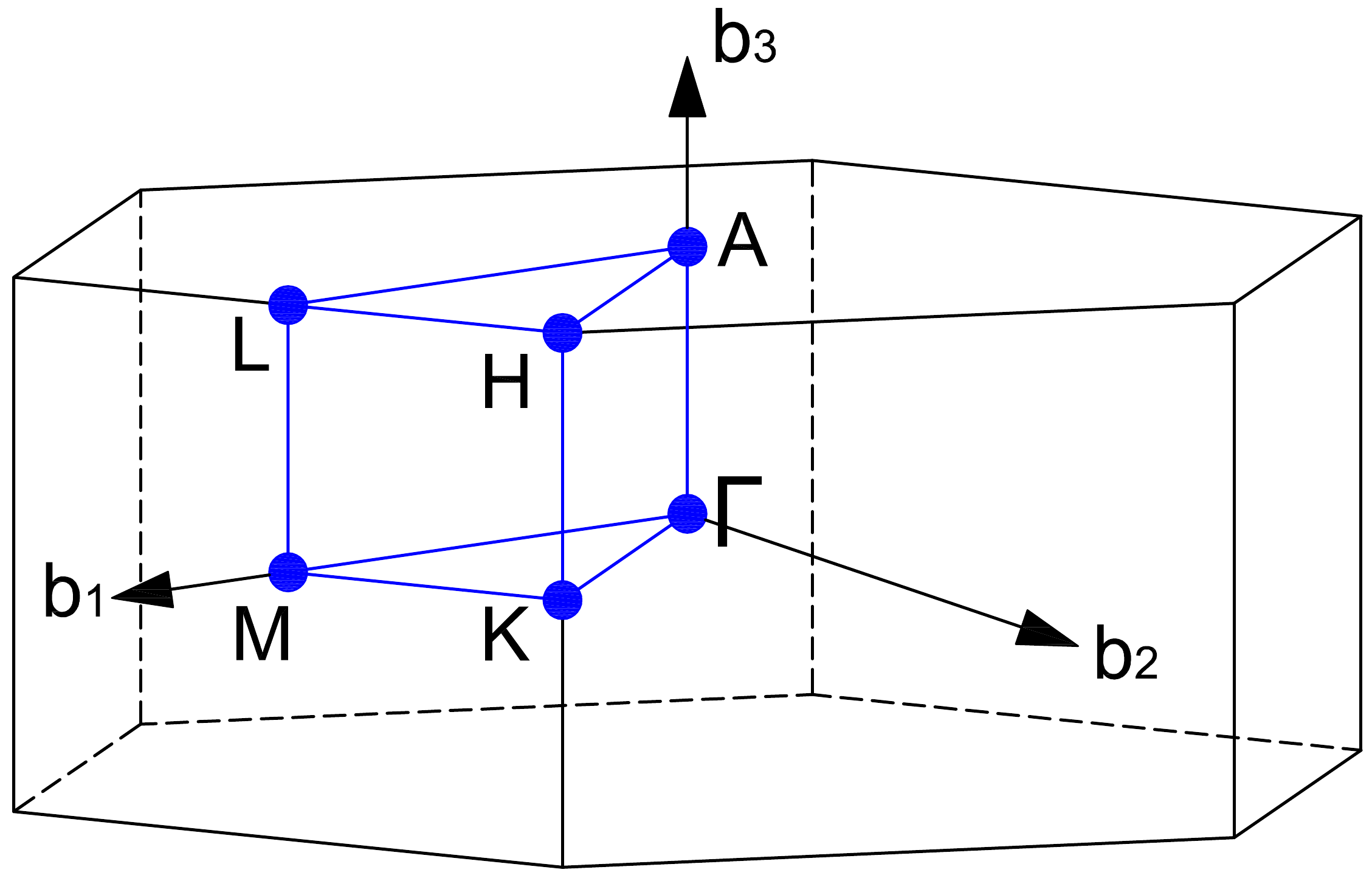}\caption{High symmetry path in the hexagonal Brillouin zone chosen according
to the notation of Ref.~\onlinecite{Setyawan10}.\label{fig0}}
\end{figure}

\begin{table}[t]
\noindent \begin{centering}
\begin{tabular}{|l||l|}
\hline 
\textbf{set} & \textbf{primitive cells} $(n_{1},n_{2},n_{3})$\tabularnewline
\hline 
\hline 
A & $(0,0,0)$, $(0,0,1)$, $(0,0,-1)$, $(0,0,2)$, $(0,0,-2)$\tabularnewline
\hline 
\multirow{2}{*}{B} & set A $\oplus$ $(1,0,0)$, $(-1,0,0)$, $(0,1,0)$, $(0,-1,0)$,\tabularnewline
 & $(1,-1,0)$, $(-1,1,0)$, $(1,1,0)$, $(-1,-1,0)$\tabularnewline
\hline 
\end{tabular}
\par\end{centering}
\caption{Sets of primitive cells used in the calculations and in the figures.\label{tab1}}
\end{table}

\begin{figure}[t]
\noindent \centering{}\includegraphics[width=1\columnwidth]{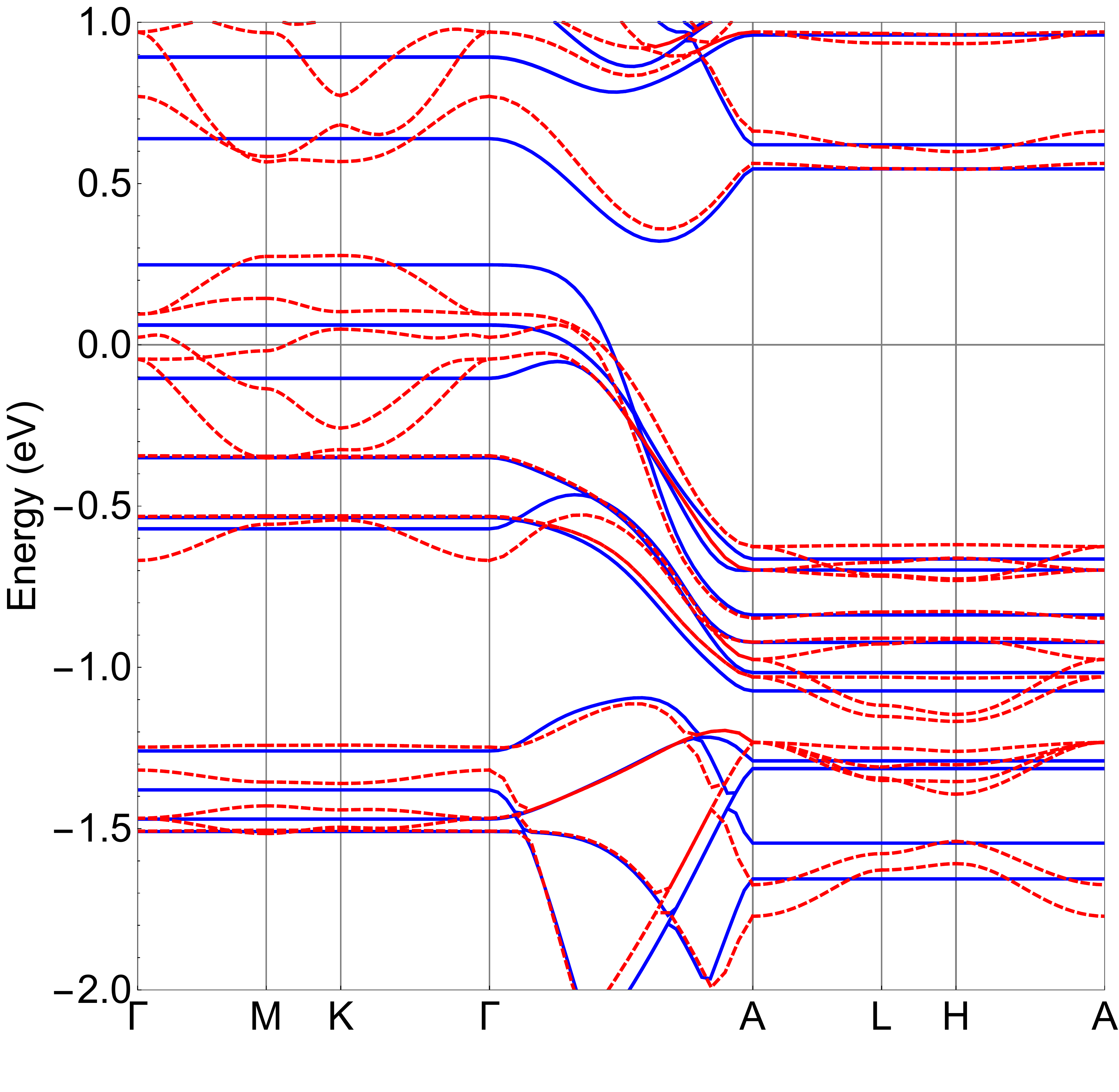}\caption{Comparison between the tight-binding band structure (blue lines),
obtained considering the primitive cells in set A of Tab.~\ref{tab1}
and the DFT spectrum (red dashed lines). The zero of energy marks
the position of the Fermi level.\label{fig1}}
\end{figure}

\begin{figure}[!t]
\noindent \centering{}\includegraphics[width=1\columnwidth]{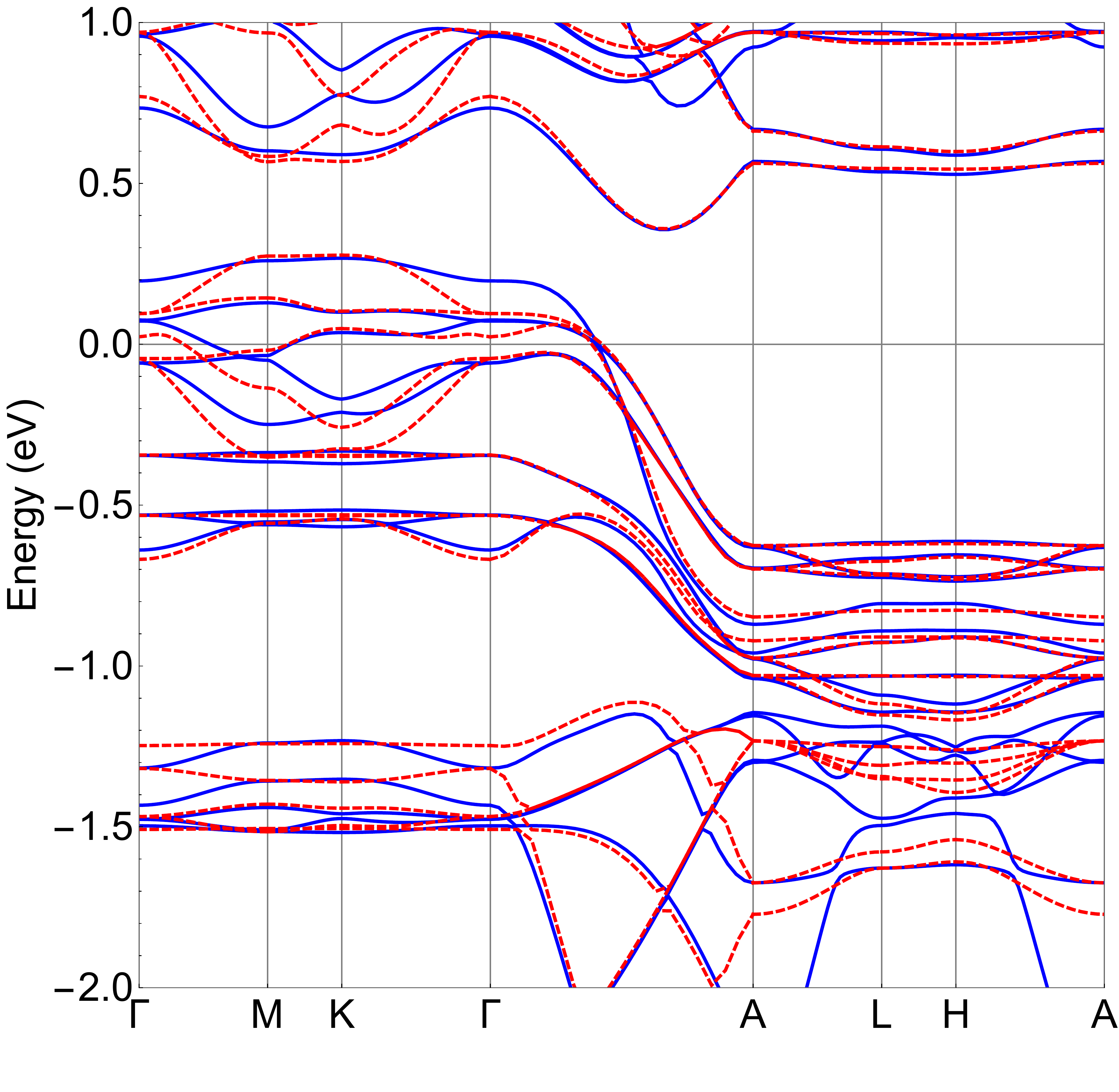}\caption{Comparison between the tight-binding band structure (blue lines),
obtained considering the primitive cells in set B of Tab.~\ref{tab1},
and the DFT spectrum (red dashed lines). The zero of energy marks
the position of the Fermi level.\label{fig2}}
\end{figure}

\section{Band structure}

To study the electronic properties of the material, we start from
the following tight-binding Hamiltonian
\begin{equation}
H=\sum_{i,\alpha,\sigma}\varepsilon_{i}^{\alpha}c_{i\alpha\sigma}^{\dagger}c_{i\alpha\sigma}-\sum_{i,j,\alpha,\beta,\sigma}t_{ij}^{\alpha\beta}(c_{i\alpha\sigma}^{\dagger}c_{j\beta\sigma}+h.c.),\label{eqn:tightbinding}
\end{equation}
in which $i$ and $j$ indicate the positions of Cr or As atoms in
the crystal, $\sigma$ is the spin, $\alpha$ and $\beta$ are the
orbital indices. The two terms take into account the on-site energies
and the hopping processes between Cr and As orbitals, respectively.
The Hamiltonian corresponds to a $48\times48$ matrix because we have
to consider the $d$ orbitals of the Chromium and the $p$ orbitals
of the Arsenic and the primitive cell contains six Cr and six As atoms.
As we have already discussed in the introduction, the hopping parameters
are obtained by the DFT overlap integrals of the orbitals. Diagonalizing
the Hamiltonian, we obtain the corresponding band structure, whose
behavior will be discussed referring to the high symmetry path in
the hexagonal Brillouin zone chosen according to the notation of Ref.~\onlinecite{Setyawan10}
(see Fig.~\ref{fig0}). In what follows, we use the notation $\bm{R}=n_{1}\bm{a}_{1}+n_{2}\bm{a}_{2}+n_{3}\bm{a}_{3}$
where $\mathbf{R}$ is a lattice vector in the direct space.

We analyze two cases that differ only for the chosen set of hopping
parameters with the aim of assessing the degree of quasi-one dimensionality
of the material. In the first case, we take into account only the
hoppings between the orbitals of the atoms that lie within a single
sub-nanotube extending along the $z$ axis. Accordingly, we limit
ourselves to consider the hoppings between the primitive cells in
set A of Tab.~\ref{tab1}. In Fig.~\ref{fig1}, we report the comparison
between the band structure obtained in this case and the DFT spectrum.
We can clearly see that this choice of hoppings not only misses completely
the in-plane band structure, but it does not even allow to describe
correctly the band structure along the $z$ axis. This unveils an
overlooked relevance of in-plane virtual processes that affect the
$z$-axis physics.

Accordingly, in the second case we have also taken into account the
in-plane hoppings between primitive cells in set B of Tab.~\ref{tab1}.
In Fig.~\ref{fig2}, we show how the agreement improves substantially
not only for the in-plane paths, which was expected, but also for
the $\Gamma$-A line of the Brillouin zone, that is for the $k_{z}$
direction. This demonstrates that it is necessary to take into account
not only the $z$-axis hopping processes, but also several in-plane
ones to get a qualitatively good description of the band structure.

In the next section, we will check whether this reduced set of hoppings
(set B) is also sufficient to obtain a qualitatively correct description,
that is in agreement with the DFT results, of the orbital populations
close to the Fermi surface.

\begin{figure}
\noindent \centering{}\includegraphics[width=1\columnwidth]{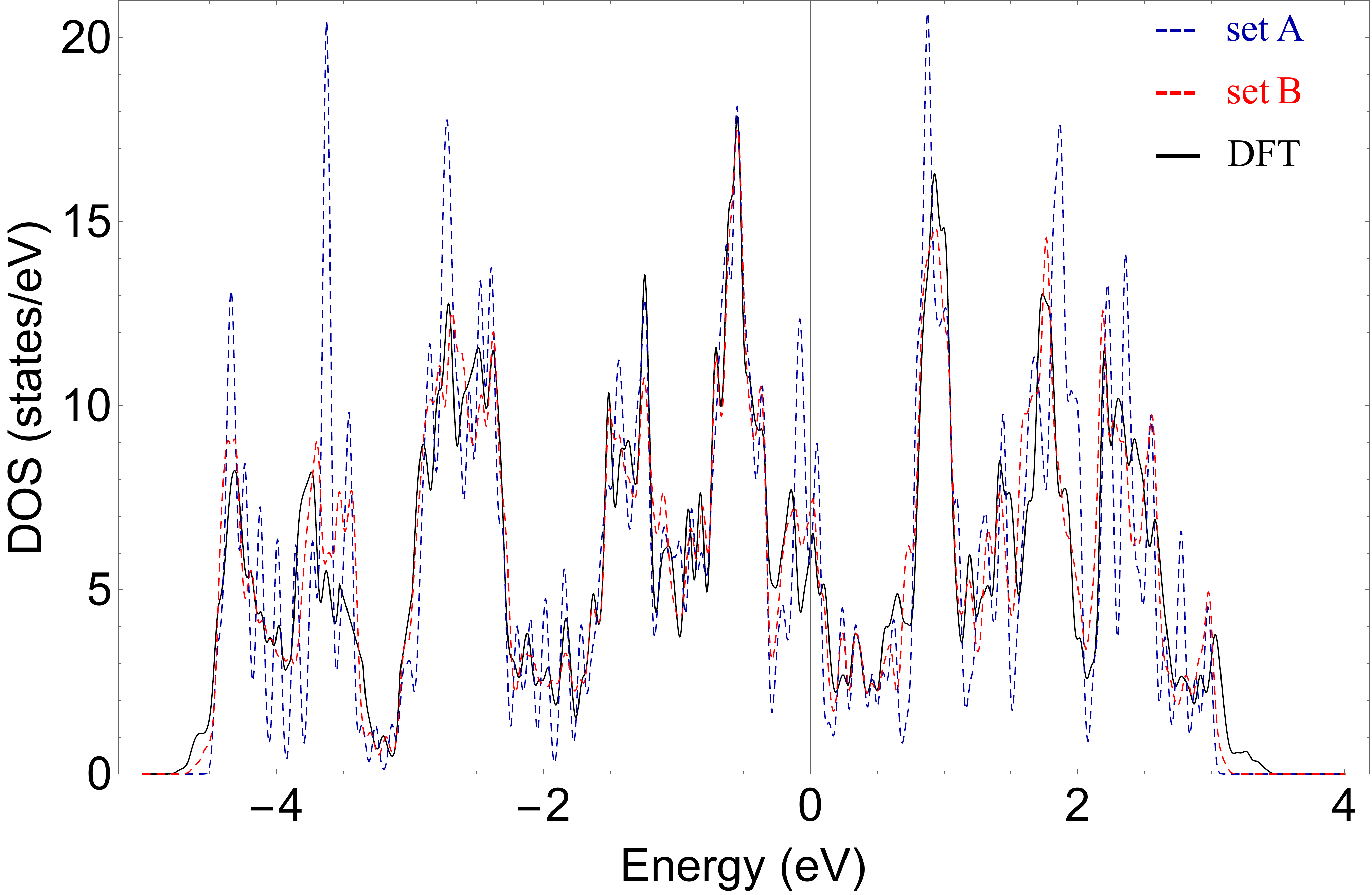}\caption{Total DOS obtained considering the primitive cells in set A and B
of Tab.~\ref{tab1} and the DFT results. The zero of energy marks
the position of the Fermi level.\label{fig3}}
\end{figure}

\begin{table}
\noindent \begin{centering}
\begin{tabular}{|c||r|r||r|r||r|r|}
\hline 
$\boldsymbol{\nu}$ & \textbf{$\boldsymbol{\rho_{\nu}^{A}\left(0\right)}$} & \textbf{$\boldsymbol{\bar{\rho}_{\nu}^{A}\left(0\right)}$} & \textbf{$\boldsymbol{\rho_{\nu}^{B}\left(0\right)}$} & \textbf{$\boldsymbol{\bar{\rho}_{\nu}^{B}\left(0\right)}$} & \textbf{$\boldsymbol{\rho_{\nu}^{C}\left(0\right)}$} & \textbf{$\boldsymbol{\bar{\rho}_{\nu}^{C}\left(0\right)}$}\tabularnewline
\hline 
\hline 
1 & 0.255 & $52.67\%$ & 0.349 & $58.51\%$ & 0.306 & $59.24\%$\tabularnewline
\hline 
2 & 0.162 & $16.73\%$ & 0.127 & $10.65\%$ & 0.116 & $11.23\%$\tabularnewline
\hline 
3 & 0.017 & $\phantom{0}3.47\%$ & 0.028 & $\phantom{0}4.69\%$ & 0.022 & $\phantom{0}4.26\%$\tabularnewline
\hline 
4 & 0.087 & $27.12\%$ & 0.104 & $26.15\%$ & 0.087 & $25.27\%$\tabularnewline
\hline 
\end{tabular}
\par\end{centering}
\caption{DOS at the chemical potential $\rho_{\nu}^{\#}\left(0\right)$ (states/eV)
and related contribution to the total DOS at the chemical potential
$\bar{\rho}_{\nu}^{\#}\left(0\right)=a_{\nu}b_{\nu}\rho_{\nu}^{\#}\left(0\right)/\rho^{\#}\left(0\right)$
for the different classes of orbitals ($\nu=1$: Cr $d_{xy},d_{x^{2}-y^{2}}$,
$\nu=2$: Cr $d_{z^{2}}$, $\nu=3$: Cr $d_{yz},d_{xz}$, 4: As $p_{x},p_{y},p_{z}$)
present in the system and different sets of hoppings $\#=A,B,C$ where
$C$ stands for the DFT results. $a_{\nu}=\left(2,1,2,3\right)$ is
the number of orbitals in the class, while $b_{\nu}=6$ is the number
of related atoms in the unit cell.\label{tab2}}
\end{table}

\section{Density of states}

\noindent We obtain the total DOS from the usual definition:
\begin{equation}
\rho(\omega)=\frac{1}{N}\sum_{\boldsymbol{k}}\delta(\omega-\varepsilon_{\boldsymbol{k}})\label{eqn:DOS}
\end{equation}

\noindent in which $\varepsilon_{\boldsymbol{k}}$ is the energy dispersion
of Hamiltonian (\ref{eqn:tightbinding}) and the sum is carried out
on the N values of momenta $\boldsymbol{k}$ in the Brillouin zone;
our numerical grid contains $6\times6\times12$ $\mathbf{k}$-points.
The delta functions in Eq.~(\ref{eqn:DOS}) have been approximated
by Gaussian functions with a variance of $\unit[0.025]{eV}$.

In Fig.~\ref{fig3}, we report the total DOS obtained taking hopping
parameters related to set A and B of Tab.~\ref{tab1}, together with
the DFT results: set B leads to a very good qualitative agreement
with the latter, this showing that this set of hoppings manages to
catch correctly also the overall populations close to the Fermi level,
in contrast to set A, which exhibits evident limitations.

We can now check, by projecting the total DOS onto the Cr and As orbitals,
if we obtain a correct description of the local orbital occupations
close to the Fermi level. To this purpose, we have reported in Tab.~\ref{tab2}
the values of the projected densities of states (PDOSs) at the Fermi
level, and its contribution to the total DOS, related to (i) the Cr-$d$
symmetrical orbitals with respect to the basal plane ($d_{xy},d_{x^{2}-y^{2}},d_{z^{2}}$),
(ii) the Cr-$d$ antisymmetrical orbitals ($d_{yz},d_{xz}$) and (iii)
the As-$p$ orbitals, for the previous three cases: set A, set B and
DFT. We notice that the value for each class, labelled by $\nu$,
has been obtained by averaging over all $a_{\nu}=\left(2,1,2,3\right)$
orbitals and $b_{\nu}=\left(6,6,6,6\right)$ atoms in the unit cell
belonging to the class. We clearly find that the symmetrical orbitals
with respect to the basal plane dominate, in agreement to what reported
in Ref.~\onlinecite{Jiang15}, and that the values found for set
B are evidently much closer to those coming from DFT (all discrepancies
below $1\%$) with respect to those obtained for set A (discrepancies
up to $6.5\%$). We conclude that set B is the minimal number of hoppings
necessary to achieve a proper description of the band structure along
the $k_{z}$ direction as well as of its orbital occupations.

It is worth noting that, as already reported in the literature~\citep{Jiang15},
the orbital characterization of the spectrum reveals a high degree
of covalency between the Cr-$d$ and the As-$p$ orbitals, as it happens
for the CrAs~\citep{Autieri17,Autieri17a,Autieri18}. This opens
up a new quest for the minimal number of effective bands necessary
to describe qualitatively, but also quantitatively, the system, at
least close to the Fermi level. We are currently investigating this
aspect too.

\section{Conclusions}

\noindent We have used a tight-binding model, obtained through the
Wannier projection of\emph{ }\textit{\emph{DFT}}\emph{ }results, to
study the electronic properties of K\textsubscript{2}Cr\textsubscript{3}As\textsubscript{3}
in order to explore its degree of unidimensionality. We have then
analyzed the band structure and the total and the local DOS of the
compound taking into account not only the usually considered hoppings
along the $z$-axis, but also those up to the next-nearest-neighbor
primitive cells in the plane. Only adding these latter, we managed
to achieve a good qualitative agreement with the DFT spectrum near
the Fermi level, demonstrating that the system does not exhibit a
genuine one-dimensional character. Consequently, this study gives
clear hints on the minimal number of hopping parameters needed to
perform further studies and understand the role of correlations, magnetic
interactions and spin-orbit as well as the mechanism and the nature
of the superconductivity in this material. A more detailed analysis
of the electronic properties of the compound, aiming to derive a reliable
minimal effective band model, is underway.
\begin{acknowledgments}
This work was supported by CNR-SPIN via the Seed Project CAMEO.
\end{acknowledgments}


%

\end{document}